\documentstyle[prl,aps]{revtex} 
\begin{document}
\draft
\twocolumn[\hsize\textwidth\columnwidth\hsize\csname
@twocolumnfalse\endcsname

\title{Decoherence within a single atom}
\author{Gyula Bene$^1$ and Szabolcs Bors\'anyi$^2$}
\address{$^1$Institute for Theoretical Physics, E\"otv\"os University,
     P\'azm\'any P\'eter s\'et\'any 1/A, H-1117 Budapest, Hungary\\
$^2$Department of Atomic Physics, E\"otv\"os University,
     P\'azm\'any P\'eter s\'et\'any 1/A, H-1117 Budapest, Hungary}
\date{\today}
\maketitle

\begin{abstract} 
 An "almost diagonal" reduced density matrix (in coordinate representation)  
 is usually a result of environment induced decoherence and is considered 
 the sign of classical behavior. We point out that the proton 
 of a ground state hydrogen atom can possess such a density matrix. 
 This example demonstrates that the "almost diagonal" structure 
 may derive from an interaction with a low number of degrees of freedom  
 which play the role of the environment. We also show that decoherence
 effects in our example can only be observed if the interaction  
 with the measuring device is significantly faster than  
 the interaction with the environment (the electron). In the opposite
 case, when the interaction with the environment is significant
 during the measurement process, coherence is maintained. Finally, 
 we propose a neutron scattering experiment on cold He atoms to
 observe decoherence which shows up as an additional positive contribution  
 to the differential scattering cross section. This contribution is inversely
 proportional to the bombarding energy.
\end{abstract} 

\pacs{03.65.-w, 03.75.Be}

\vskip2pc]
\narrowtext 
Decoherence\cite{decoherence} is a remarkable quantum phenomenon which plays a central 
role in the (not yet fully understood) emergence of classical 
properties. It is also a practically important effect in  
quantum optics, atomic physics and mesoscopic systems. Therefore,  
it can be of interest to discuss some of its less known aspects. 
In the present paper we show on a simple example that 
contrary to the common belief, 
\begin{enumerate} 
\item an "almost diagonal" (in coordinate representation) 
density matrix may be the result of an interaction with only 
a few degrees of freedom 
\item an "almost diagonal" density matrix does not necessarily 
imply the loss of coherence 
\item interaction may maintain rather than destroy coherence.  
\end{enumerate}     
We also discuss the conditions when and how decoherence can be 
observed.  
 
Our primary example is a single ground state hydrogen atom. Later we shall 
also consider the case of a helium atom.   

The wave function of a freely moving hydrogen atom in the ground state 
may be written in coordinate representation in the product form 
\begin{eqnarray} 
\Phi(\vec{r}_p,\vec{r}_e)=\psi(\vec{R})\varphi_0(\vec{r})\label{st1} 
\end{eqnarray} 
where 
\begin{eqnarray} 
\vec{R}=\frac{m_p\vec{r}_p+m_e\vec{r}_e}{m_p+m_e} 
\end{eqnarray} 
stands for the coordinates of the center of mass and 
\begin{eqnarray} 
\vec{r}=\vec{r}_e-\vec{r}_p 
\end{eqnarray} 
for the relative coordinates. The vectors  
$\vec{r}_p$, $\vec{r}_e$ refer to the position 
of the proton and that of the electron, respectively, while  
$m_p$ and $m_e$ are the corresponding masses. 
In Eq.(\ref{st1})  
\begin{eqnarray} 
\varphi_0(\vec{r})=\frac{1}{\sqrt{\pi a_B^3}}\exp\left(-\frac{r}{a_B}\right) 
\end{eqnarray} 
is the ground state of the hydrogen atom, where  
$a_B=\frac{\hbar^2}{m_e e^2}=0.529\times 10^{-8} {\rm cm}$ stands for the  
Bohr radius. 
As the Schr\"odinger equation of the hydrogen atom separates 
in the $\vec{R}$, $\vec{r}$ coordinates,  
the time evolution of the center of mass wave function 
does not influence the state $\varphi_0(\vec{r})$, and the 
product form (\ref{st1}) persists. 
Nevertheless, when expressing the total wave function in the original 
coordinates $\vec{r}_p$, $\vec{r}_e$, the state of the hydrogen atom 
is not of a product form, instead, it is an entangled state. 
Indeed, both $\psi(\vec{R})$ and $\varphi_0(\vec{r})$ depend  
on both $\vec{r}_p$ and $\vec{r}_e$. For simplicity suppose 
that $\Delta P\;a_B/\hbar \ll m_p/m_e$ ($\Delta P$ standing for the  
momentum uncertainty of the center of mass), or, equivalently, that 
\begin{eqnarray}
\Delta V\ll\frac{\hbar}{m_e a_B}\approx v_e \approx2\times 10^6 \frac{m}{s}\label{cond}
\end{eqnarray}
($\Delta V$ being the velocity uncertainty of the center of mass and
$v_e$ standing for the velocity of the electron), then $\psi(\vec{R})\approx 
\psi(\vec{r}_p)$, and we arrive at the Born-Oppenheimer approximation\cite{Born-Opp} 
\begin{eqnarray} 
\Phi(\vec{r}_p, \vec{r}_e)=\psi(\vec{r}_p)\varphi_0(\vec{r}_e-\vec{r}_p) 
\label{st2} 
\end{eqnarray}   
Entanglement implies that the proton 
is not in a pure state, i.e., its density matrix $\rho(\vec{r}_p, \vec{r}_p{\mbox{\hspace{0.1mm}}'})$ is not 
a single projector. Indeed, using expression (\ref{st2}) we get 
\begin{eqnarray} 
\rho(\vec{r}_p, \vec{r}_p{\mbox{\hspace{0.1mm}}'})=\int d^3\vec{r}_e  
\Phi(\vec{r}_p, \vec{r}_e)\Phi^*(\vec{r}_p{\mbox{\hspace{0.1mm}}'}, \vec{r}_e)\label{ro1}\\ 
=\psi(\vec{r}_p)\psi^*(\vec{r}_p{\mbox{\hspace{0.1mm}}'})\left(1+s+\frac{1}{3}s^2\right){\rm e}^{-s}\label{ro2} 
\end{eqnarray} 
where $s=|\vec{r}_p-\vec{r}_p{\mbox{\hspace{0.1mm}}'}|/a_B$. 
Provided that the width of the center of mass wave function 
is much larger than the Bohr radius, the reduced density matrix  
(\ref{ro2}) is "almost 
diagonal", i.e., its width (which is approximately $2\;a_B$) is much 
smaller than its length along the diagonal. The condition 
for this is $\Delta P\;a_B/\hbar\ll 1$, or  
\begin{eqnarray}
\Delta V\ll\frac{\hbar}{m_p a_B}\approx 10^3 \frac{m}{s}\;.\label{dec_cond} 
\end{eqnarray} 
 It is a much stronger condition than that of the 
Born-Oppenheimer approximation (cf. Eq.(\ref{cond})). 
Let us emphasize that the "almost diagonal" structure in this example
is due to the interaction with a single electron.
  
 An "almost diagonal" density matrix (in coordinate representation)  
 is usually  
 a result of "environment induced decoherence" and is considered 
 the sign of classical behavior.  
In the present paper we  
consider the validity of this expectation and discuss  
the physical meaning of density matrices like (\ref{ro2}). 
 
It is of worth mentioning that we do not need to rely upon 
the Born-Oppenheimer approximation in order to arrive at an "almost diagonal" density matrix. 
That approximation 
is only convenient because it allows one to perform the integration  
in Eq.(\ref{ro1}) explicitly, independently of the 
actual form of the center of mass wave function.  
More generally, Eqs.(\ref{st1}), (\ref{ro1}) imply 
\begin{eqnarray}
\left|\rho_p(\vec{r}_p, \vec{r}_p{\mbox{\hspace{0.1mm}}'})\right|\le \left(\max_{\vec{R}} \left|
\psi(\vec{R})\right|^2\right)\left(1+s+\frac{1}{3}s^2\right){\rm e}^{-s}\label{ro3}
\end{eqnarray}
This inequality shows that the offdiagonal matrix elements of the reduced density
matrix decay at least exponentially when increasing $s=|\vec{r}_p-\vec{r}_p{\mbox{\hspace{0.1mm}}'}|/a_B$. 
 
As an example, suppose that 
\begin{eqnarray} 
\psi(\vec{R})=\frac{1}{(2\pi\delta^2)^{\frac{3}{4}}}\frac{1}{\left(1+i\frac{\hbar t}{2 M \delta^2}\right)^{\frac{3}{2}}}
\exp\left(-i\frac{P_0^2 t}{2 M \hbar}\right.\nonumber\\\left.
-\frac{(\vec{R}-\vec{R}_0-\frac{\vec{P}_0}{M}t)^2}{4\delta^2\left(1+i\frac{\hbar t}{2 M \delta^2}\right)}
+i\frac{\vec{P}_0\cdot (\vec{R}-\vec{R}_0)}{\hbar}\right)\label{gauss}
\end{eqnarray}  
where $M=m_p+m_e\approx m_p$. Eq. (\ref{gauss}) describes a Gaussian wave packet moving in free space.  
If condition (\ref{cond}) does not hold, the integral in Eq.(\ref{ro1}) cannot be expressed 
in a closed analytical form. Assuming the validity of (\ref{cond}), however, we obtain Eq.(\ref{ro2}).
We get a narrow density matrix (an "almost diagonal" one) if 
$\Delta x=\sqrt{\delta^2+\left(\frac{\hbar t}{2 M\delta}\right)^2}\gg a_B$. 

The deviation of the reduced density matrix of the proton  from the pure state can be characterized
by ${\rm Tr} \hat \rho_p^2$. It is unity for a pure state and less than unity (but positive)
for a mixed state. Using Eqs.(\ref{gauss}), (\ref{cond}) we have
\begin{eqnarray} 
{\rm Tr} \hat \rho_p^2=\frac{z^3}{2\sqrt{\pi}}\int_0^\infty ds\;s^2\left(1+s+\frac{1}{3}s^2\right)^2\nonumber\\
\times \exp\left(-2s-\frac{s^2 z^2}{4}\right)
\label{sp1} 
\end{eqnarray}   
where 
\begin{eqnarray} 
z=\frac{a_B}{\Delta x}=\frac{a_B}{\sqrt{\delta^2+\left(\frac{\hbar t}{2 M\delta}\right)^2}}
\end{eqnarray} 
It is easy to see that in case of an "almost diagonal" density matrix
(i.e., when $z\ll 1$)
${\rm Tr} \hat \rho_p^2$ is much smaller than one (in fact, it goes to zero with $z$ as $z^3$),
while in the other extreme of very large $z$ it approaches unity, which corresponds to a pure state.

As it is well known, a reduced density matrix gives all the possible information
about the results of measurements done on the corresponding subsystem.
Let us emphasize that such measurements are assumed to be infinitely fast, so
that they give a snapshot about the state of the subsystem. It is obvious that
coordinate measurements do not reveal the structure of the density matrix,
as the probability distribution of the coordinates are given by the
diagonal elements $\rho_p(\vec{r}_p, \vec{r}_p)$ which is independent of
the decay of the offdiagonal elements. Note that in the Born-Oppenheimer
approximation (cf. Eq.(\ref{st2})) $\rho_p(\vec{r}_p, \vec{r}_p)=|\psi(\vec{r}_p)|^2$,
which is completely independent of the interaction between the proton and the electron.
In contrast, momentum measurements are sensitive for the behavior of the offdiagonal elements
of the reduced density matrix in coordinate respresentation.
The probability distribution of the momentum is determined by the diagonal elements
of the reduced density matrix in momentum representation, i.e., by 
\begin{eqnarray} 
\tilde \rho_p(\vec{p}_p, \vec{p}_p)
=\frac{1}{(2\pi \hbar)^3}\int d^3 \vec{r}_p 
\int d^3 \vec{r}_p{\mbox{\hspace{0.1mm}}'}\;\rho_p(\vec{r}_p,
\vec{r}_p{\mbox{\hspace{0.1mm}}'})
\nonumber\\\times
\exp\left(-\frac{i}{\hbar}\vec{p}_p\cdot (\vec{r}_p-\vec{r}_p{\mbox{\hspace{0.1mm}}'})\right)\label{rop1} 
\end{eqnarray} 
Introducing coordinates parallel and perpendicular to the diagonal of the reduced 
density matrix (in coordinate representation), i.e.,
\begin{eqnarray}
\vec{r}_{\parallel}=\frac{1}{2}(\vec{r}_p+\vec{r}_p{\mbox{\hspace{0.1mm}}'})\\
\vec{r}_{\perp}=\vec{r}_p-\vec{r}_p{\mbox{\hspace{0.1mm}}'}
\end{eqnarray}  
we may write
\begin{eqnarray} 
\tilde \rho_p(\vec{p}_p, \vec{p}_p)=\frac{1}{(2\pi \hbar)^3}
\int d^3 \vec{r}_{\perp}
\exp\left(-\frac{i}{\hbar}\vec{p}_p\cdot \vec{r}_{\perp})\right)\nonumber\\\times
\int d^3 \vec{r}_{\parallel}\; \rho_p(\vec{r}_{\parallel}+\frac{1}{2}\vec{r}_{\perp},\; \vec{r}_{\parallel}-\frac{1}{2}\vec{r}_{\perp})
\label{rop2} 
\end{eqnarray}
This last expression shows that the momentum distribution is proportional to the Fourier transform
of the offdiagonal elements of the density matrix in coordinate representation, after having
averaged them along the diagonal. Obviously, the narrower the density matrix (in coordinate representation)
becomes, the broader the momentum distribution will be. Using the Born-Oppenheimer approximation (\ref{st2}) and 
the explicit expression (\ref{gauss}), we get 
\begin{eqnarray} 
\tilde \rho_p(\vec{p}_p, \vec{p}_p)=\frac{a_B^3}{2\pi^2\hbar^3}\frac{1}{q}\int_0^\infty ds\; s\; \sin(qs)\nonumber\\
\times\left(1+s+\frac{1}{3}s^2\right)
 \exp\left(-s-\frac{s^2 z_0^2}{8}\right) 
\label{rop3} 
\end{eqnarray}
where $q=|\vec{p}_p-\vec{P}_0|a_B/\hbar$ and $z_0=a_B/\delta$. Note that the momentum distribution is
independent of the time, which is a natural consequence of the fact that momentum
(unlike coordinates) is now a conserved quantity.      
 
It is instructive to evaluate expression (\ref{rop3}) in two extreme 
situations: for $a_B\gg \delta$ and $a_B\ll \delta$.  
The former case corresponds to a pure state of the proton, while in the 
latter case  
one has an "almost diagonal" mixed state. For $a_B\gg \delta$ 
we get 
\begin{eqnarray} 
\tilde \rho_p(\vec{p}_p, \vec{p}_p)=\left(\frac{2}{\pi}\right)^{\frac{3}{2}}\frac{\delta^3}{\hbar^3} 
\exp\left(-2\;\frac{(\vec{p}-\vec{P_0})^2\delta^2}{\hbar^2}\right)\label{pdist1} 
\end{eqnarray} 
It coincides with the momentum distribution of the center of mass, 
and is independent of the electronic motion. In the opposite case 
(which is our case of interest), for $a_B\ll \delta$ we have 
\begin{eqnarray} 
\tilde \rho_p(\vec{p}_p, \vec{p}_p)=\frac{8a_B^3}{\pi^2 \hbar^3}\frac{1}{(1+q^2)^4}\label{pdist2} 
\end{eqnarray}    
This coincides with the momentum distribution of the atomic electron (if $\vec{q}$ is 
identified with the momentum of the electron times $a_B/\hbar$) 
and is independent of the center of mass motion. Let us emphasize 
that it is the momentum distribution of the proton, and the above coincidence  
is a consequence of momentum conservation and that the variance  
of the total momentum is negligible. Obviously, the width of distribution 
(\ref{pdist2}) is $\approx \hbar/a_B$, much larger than $\hbar/\delta$, 
the momentum uncertainty corresponding to the state (\ref{gauss}).  
This broadening marks the ``almost diagonal'' structure of 
$\rho(\vec{r}_p, \vec{r}_p{\mbox{\hspace{0.1mm}}'})$.  
Below we shall point out that this sign of decoherence also shows up in 
neutron scattering cross sections which are experimentally more accessible. 
 
The above discussion  
raises the obvious question how one can then observe diffraction 
with atomic beams \cite{atom_interference}, why interference 
is not destroyed by decoherence. In a typical diffraction experiment 
done with atoms the condition of ``almost diagonal'' structure of 
$\rho(\vec{r}_p, \vec{r}_p{\mbox{\hspace{0.1mm}}'})$, Eq.(\ref{dec_cond}) 
is indeed fulfilled. The essential point of the explanation is that the 
density matrix which characterizes a subsystem (in our case the proton) 
refers to instantaneous (infinitely fast) measurements. This means  
that one can observe decoherence effects in a measurement done on the 
proton if during the interaction between the proton and the  
measuring device the state of the electron does not change  
significantly. As an estimate, one may compare the momentum actually
gained by the electron from the proton during the measurement 
to the momentum change of the electron that would take place 
if the electron followed the proton instantaneously (as in the 
Born-Oppenheimer approximation). 
One can also tell that decoherence effects become observable if
the measurement is so fast that the Born-Oppenheimer approximation
for the electronic motion fails. 
Later we shall express the condition in a concrete situation quantitatively.  
The condition of observing decoherence is obviously 
not fulfilled in a diffraction experiment: the time of the measurement 
here is the time of flight between the grid and the screen.  
During this time the interaction between the proton and the electron 
 maintains the form (\ref{st1}) of the state  
of the atom, i.e., the Born-Oppenheimer approximation is valid all the
time. For definiteness, let us consider a two slit  
experiment where the distance between the slits is much larger than
$a_B$ but smaller than $\delta$. Be $\alpha(\vec{R},t)$ and $\beta(\vec{R},t)$  
outgoing waves emerging from the 
first and the second slit, respectively. Then the center of mass 
wave function can be written as  
\begin{eqnarray} 
\psi(\vec{R},t)=a\alpha(\vec{R},t)+b\beta(\vec{R},t)\;,\label{slit1} 
\end{eqnarray} 
where $a$ and $b$ are the probability amplitudes that the atom 
goes through the first and the second slit, respectively. 
Eq.(\ref{st1}) remains valid until the atom reaches the screen, thus 
Eq.(\ref{slit1}) is also valid during the time of flight. Initially, 
at the instant of time when the atom has passed through the slits 
$\alpha(\vec{R},t)$ and $\beta(\vec{R},t)$ are still narrow, separated 
wave packets, but later on they broaden and overlap. At the (only approximately 
determined) time instant $t_0$ when the atom hits the screen  
Eqs.(\ref{st1})-(\ref{ro2}) are still valid, thus we get for the  
probability distribution $P(\vec{r}_p)$ of finding the proton near  
a given point $\vec{r}_p$ the expression 
\begin{eqnarray} 
P(\vec{r}_p)=|a\alpha(\vec{r}_p,t_0)+b\beta(\vec{r}_p,t_0)|^2\;.\label{slit2} 
\end{eqnarray}             
This includes interference terms, too. Thus we see that despite of 
the "almost diagonal" density matrix (which has had such form 
already initially) coherence is not destroyed. 
Paradoxically, this is due to the same interaction (Coulomb attraction 
between the electron and the proton) which is 
responsible for the "almost diagonal" structure of the density matrix. 
 
In order to shed more light on the situation, let us consider 
what we would get if the interaction between the electron and the  
proton were turned off when the atom had passed through the slits.  
This can be done in principle even experimentally by making use of the
fact that during fast electronic processes the nucleus remains
``frozen''. In molecular physics this is called the
Franck-Condon principle\cite{Franck-Condon}. If the atom is irradiated by 
a suitable ultraviolet laser beam, it can be ionized. This is a fast 
process which practically  
does not influence the position and motion  of the proton, but eliminates 
the interaction between the proton and the electron.  
In order to see the consequences let us express 
the initial state $\Phi(\vec{r}_p, \vec{r}_e,t=0)$ (i.e., still before
the ionization) in the Schmidt  
(or biorthogonal) representation\cite{Schmidt}, 
i.e., as 
\begin{eqnarray} 
\Phi(\vec{r}_p, \vec{r}_e,t=0)=\sum_j c_j \chi_j(\vec{r}_p)
\xi_j(\vec{r}_e)\nonumber \\ 
\approx a\alpha(\vec{r}_p,t=0)\varphi_0(\vec{r}_e-\vec{r}_{s1})\nonumber\\ 
+b\beta(\vec{r}_p,t=0)\varphi_0(\vec{r}_e-\vec{r}_{s2})\;.\label{sch} 
\end{eqnarray}  
Here $\vec{r}_{s1}$, $\vec{r}_{s2}$ stand for the slit positions, 
$\chi_j(\vec{r}_p)$-s (i.e., $\alpha(\vec{r}_p,t=0)$ and
$\beta(\vec{r}_p,t=0)$) are the eigenstates of the reduced density 
matrix of the proton, while $\xi_j(\vec{r}_e)$-s (i.e.,
$\varphi_0(\vec{r}_e-\vec{r}_{s1})$
and $\varphi_0(\vec{r}_e-\vec{r}_{s2})$) are the eigenstates 
of the reduced density matrix of the electron.
The approximate orthogonality of $\varphi_0(\vec{r}_e-\vec{r}_{s1})$
and $\varphi_0(\vec{r}_e-\vec{r}_{s2})$ follows because
$|\vec{r}_{s1}-\vec{r}_{s2}|\gg a_B$.
During and after the ionization process the states of the 
electron-photon system evolve in time unitarily. The states
$\chi_j(\vec{r}_p)$ of the proton have a separate unitary
time evolution.
Therefore, we have 
\begin{eqnarray} 
P(\vec{r}_p)= 
\sum_j |c_j|^2\left|\left(\hat
    U_{t_0}\varphi_j\right)(\vec{r}_p)\right|^2\nonumber
\\ 
\approx |a|^2 |\alpha(\vec{r}_p,t_0)|^2+|b|^2|\beta(\vec{r}_p,t_0)|^2 
\;,\label{slit3} 
\end{eqnarray}   
i.e., interference terms are absent. It can also be understood 
as a consequence of the fact that the "which way" information 
after the ionization
can be obtained by measuring the  
electron without disturbing the proton.   
 
In the presence of the interaction, 
however, interference and coherence is restored. In that case, of course, 
separate unitary time evolutions for the proton and the electron  
do not exist, instead, they move together and the atom behaves (in free space or in 
slowly varying potentials) as a single unit. 
 
Let us return now to the question how one could observe experimentally 
the consequences of the "almost diagonal" density matrix (\ref{ro2}). 
As mentioned above, decoherence effects emerge if during the 
measurement done on the proton the interaction between the electron 
and the proton is negligible. One possibility, namely,  
photoionization behind the slit has already been 
mentioned. In this case 
the duration of the measurement is unchanged, but the electron-proton 
interaction is "turned off". Another possibility is to perform a very fast 
measurement on the proton. Below we suggest such an experiment. 
 
Let us consider low energy (a few eV-s) neutron scattering 
at a helium atom, prepared in a state where the width of the center of mass 
wave function is much larger than the atomic size. Low energy 
neutrons interact with the nucleus through a contact potential 
\begin{eqnarray} 
g\delta(\vec{r}_n-\vec{r}_\alpha) \label{cont_pot} 
\end{eqnarray} 
where $g=2\pi \hbar^2 a/\mu$, $a$ standing for the scattering length 
and $\mu$ being the reduced mass of the neutron-nucleus system. 
There is also an interaction between the magnetic moments 
of the neutron and the electron, however, in case of the helium atom 
the contributions of the two electrons to the magnetic scattering 
cancel each other, provided that an inelastic scattering (excitation 
of the electrons to higher levels) is energetically not possible.     
This is why we need helium for this experiment rather than hydrogen. 
 
The condition for observing decoherence is that during the interaction 
time $d/v$ ($d$ standing for the nucleus size and $v$ for the neutron 
velocity) the momentum transferred to the electron $$\approx
\frac{d}{v}\frac{1}{4\pi\epsilon_0}\frac{2 q^2}{a_B^2}\approx
\frac{m_e v_e^2}{v}\frac{d}{a_B}$$ is much less than the total change
of the momentum of the electrons well after the collision, $2 m_e v$.
(As before, $m_e$ and $v_e$ stand for the electron mass and velocity,
respectively.)
Thus the condition is
\begin{eqnarray}
v\gg \sqrt{\frac{d}{a_B}} v_e \approx 4\times 10^3 \frac{m}{s}\label{cdx}
\end{eqnarray}
In other terms, the bombarding neutron energy must be much larger than
$0.08\; eV$. This ensures that the electron-$\alpha$ interaction is negligible
during the neutron-$\alpha$ collision. 

The suggested measurement is just a measurement of the differential
scattering cross section of the neutron-helium collision (the neutron
is detected) under the conditions described above. We also assume
that the incoming neutron can be described by a plane wave in a
satisfactory manner. First order time dependent perturbation
theory gives a suitable approximation for the wave function of the
whole system, then we calculate the probability of observing an
outcoming neutron at a given angle with arbitrary momentum.
Thus we get for the differential scattering cross section the formula
(in laboratory frame)  
\begin{eqnarray} 
\frac{d\sigma}{d\Omega}
=\frac{m_n g^2}{(2\pi\hbar)^3 k}
\int_0^\infty dk' \int_{-\infty}^\infty d\tau \int d^3\vec{y}_{\parallel}  
\;k'^2 \nonumber\\\times
 \exp\left[-i
    \hbar\left(\frac{k^2-k'^2}{2m_n}
        -\frac{(\vec{k}-\vec{k}')^2}{2M_\alpha}\right)\tau\right]\nonumber\\\times
\rho\left(\vec{y}_{\parallel}  
+ \frac{\hbar}{2M_\alpha}(\vec{k}-\vec{k}')\tau,\; \vec{y}_{\parallel}  
- \frac{\hbar}{2M_\alpha}(\vec{k}-\vec{k}')\tau\right) \label{crossec1}
\end{eqnarray}  
Here $\vec{k}$, $\vec{k}'$ stand for the wave vectors of the incoming 
and the scattered neutrons, respectively. 
The explicit appearence of the density matrix of the $\alpha$ particle is
just a sign of decoherence, which in turn is a consequence of the
negligible interaction between the nucleus and the electrons during
the collision. In order to calculate the density matrix of the nucleus 
of the helium atom we again apply the Born-Oppenheimer approximation  
(\ref{st2}) where instead of the one-electron wave function 
we have a two-electron wave function. For the ground state we 
use the hidrogen-like wave function 
\begin{eqnarray} 
\varphi_0(\vec{r_1},\vec{r_2},\sigma_1,\sigma_2)= 
\frac{Z^{*3}}{\pi a_B^3}\exp\left(-\frac{Z^*}{a_B}(r_1+r_2)\right) 
\nonumber\\\times
\chi^0(\sigma_1,\sigma_2)\;.\label{Hest1} 
\end{eqnarray}  
Here  
$
Z^*=\frac{27}{16} 
$  
is the effective atomic number and 
$\chi^0(\sigma_1,\sigma_2)$ stands for the singlet spin function.  
Using Eq.(\ref{Hest1}) we get for the density matrix of the nucleus 
\begin{eqnarray} 
\rho(\vec{r}_\alpha, \vec{r}_\alpha{\mbox{\hspace{0.1mm}}'})  
=\psi(\vec{r}_\alpha)\psi^*(\vec{r}_\alpha{\mbox{\hspace{0.1mm}}'})\left(1+s+\frac{1}{3}s^2\right)^2{\rm e}^{-2s}\label{Hero2}  
\end{eqnarray}  
where $s=Z^*|\vec{r}_\alpha-\vec{r}_\alpha{\mbox{\hspace{0.1mm}}'}|/a_B$. 
 
Inserting Eqs.(\ref{Hero2}) and (\ref{gauss}) into Eq.(\ref{crossec1})  
we get (if \mbox{$P_0=0$})

\begin{eqnarray} 
\frac{d\sigma}{d\Omega}=\frac{m_n g^2}{(2\pi\hbar)^3
  k}\int_{-\infty}^\infty d\tau \int_0^\infty dk'\; 
k'^2  \nonumber\\ 
\times\;{\rm e}^{-2\kappa |\tau|}\left(1+\kappa
  |\tau|+\frac{1}{3}\kappa^2\tau^2\right)^2\nonumber\\ 
\times
\exp\left[-i 
    \hbar\left(\frac{k^2-k'^2}{2m_n} 
        -\frac{(\vec{k}-\vec{k}')^2}{2m_\alpha}\right)\tau
-\frac{1}{8}z_0^2\kappa^2\tau^2\right] 
\end{eqnarray} 
where 
\begin{eqnarray} 
\kappa=\frac{\hbar}{m_\alpha}\frac{|\vec{k}-\vec{k}'|}{a_B}Z^* 
\end{eqnarray}  

 Finally, by doing the integrations in the limit $a_B\ll \delta$ and
\begin{eqnarray} 
q=k a_B=\frac{\sqrt{2m_n E_n}}{\hbar} a_B\gg 1 
\end{eqnarray} 
(note that Eq.(\ref{cdx}) implies $q\gg 3.4$) we have (by taking
$m_\alpha=4 m_n$)
\begin{eqnarray} 
\frac{d\sigma}{d\Omega}=\frac{m_n^2 g^2}{25\pi^2\hbar^4}f(\vartheta)
\left(1+\frac{h(\vartheta)}{q^2}+{\cal O}\left(\frac1{q^4}\right)\right) \;.\label{vegeredmeny}
\end{eqnarray}

where
\begin{equation}
f(\vartheta)=\frac{\left(\cos\vartheta+\sqrt{15+\cos^2\vartheta}\right)^2}{\sqrt{15+\cos^2\vartheta}}\label{nullrend}
\end{equation}
and
\begin{eqnarray}
h(\vartheta)=\frac{6075}{64}\frac{3+5\cos^2\vartheta}
{\left(15+\cos^2\vartheta\right)^2\left(\cos\vartheta+\sqrt{15+\cos^2\vartheta}\right)^2}\;.
\end{eqnarray}
Note that Eq.(\ref{vegeredmeny}) is given in laboratory frame and the
angular dependence (\ref{nullrend}) of the leading order corresponds just to
isotropic scattering in the center of mass frame.
 Decoherence shows up as an anomalous 
contribution to the differential scattering cross section that is
inversely proportional to the bombarding energy. This contribution is
always positive and has
maxima at $\vartheta=0$ and $\vartheta=\pi$. At $E_n=1\; eV$
these maxima are  $h(0)/q^2=8.2\times10^{-4}$ and $h(\pi)/q^2=2.27\times10^{-3}$.

In conclusion we add that other kinds of interactions (e.g.,
electron-photon or gravitational) may also result in an almost diagonal
reduced density matrix. Again, one has to be careful in interpreting
it as a loss of coherence. A more detailed discussion of these
questions will be presented elsewhere.
\vskip0.5cm
{\bf  Acknowledgement}
\vskip0.5cm
One of the authors (Gy.B.) is indebted to T.Geszti
for an enlightening discussion.
 This work has been partially supported by the Hungarian Aca\-demy of
 Sciences                                         
 under Grant No. OTKA T 029752 and the J\'anos Bolyai Research Fellowship.

\end{document}